\documentclass[reqno]{article}
\usepackage{alltt}
\usepackage{graphicx}
\title{Reconstruction of the unitary symmetry in super-relativity}
\author{Peter Leifer}
\date{Cathedra of Informatics, Crimea State Engineering and
Pedagogical University, \\
21 Sevastopolskaya st., 95015 Simferopol, Crimea, Ukraine; \\Hermon Laboratories, Ltd. \\
Binyamina, 30500 Israel \\
leifer@bezeqint.net, peter@hermonlabs.com }
\begin{document}
\maketitle
\begin{abstract}
The reconstruction of the unitary symmetry \cite{TFD} under non-linear dynamical
mapping Hilbert space of action amplitudes $C^N$ onto projective Hilbert space
$CP(N-1)$ \cite{Le1} has been applied here to the quantum dynamics of elementary
vacuum excitations. The ``vacuum manifold of virtual action states" is represented here
by $CP(N-1)$ whereas its tangent vectors define local dynamical variables (LDV's)
describing ``matter". The conservation laws of LDV's express self-conservation
of the ``material particles" during continuous evolution
being expressed as the affine parallel transport agrees with Fubuni-Study metric,
create the ``affine gauge potential" as the solution of the partial differential
equations. Such procedure embeds the quantum dynamics into dynamical space-time whose
state-dependent coordinates arose due to encoding results of quantum measurement by
the qubit spinor whose components subjected to Lorentz transformations of ``quantum boosts"
and ``quantum rotations". Thereby, in the framework of super-relativity, the objective
character of the quantum measurement is inherently related to the dynamical
space-time structure that replaces the notion of ``observer".
\end{abstract}


\section{Introduction}
Space-time being macroscopically observable
as global pseudo-Riemannian manifold may emerge due to newly defined objective
quantum measurement \cite{Le2}. It means that (objective) quantum interaction may
be used as an operational procedure for ``marking" non-local quantum ``lump" used as
a ``pointer". I would like to recall that the localization and ``marking" classical events by means of {\it classical electromagnetic field} is based on the distinguishability (separability), i.e. individualization of pointwise material objects. However we loss the possibility to distinguish (non-local) quantum objects by mean of {\it quantum fields} in
space-time and, hence, it is impossible \textbf{directly} to identify them with
space-time points. Therefore obviously non-local quantum particles are too complicated
objects in order to be used as fundamental \emph{primordial} elements of quantum theory.
The pure states of quantum motion with quantized finite action used here as a fundamental element
that should be localized in some manifold. The projective Hilbert state space $CP(N-1)$
takes the place of such manifold; dynamics of non-local particles, evolution and
measurement procedure formulated in fibre bundle over this base manifold. The space-time
notion may be introduced only as a manifold of some coordinates prescribed by
self-consistent manner to ``lump". We have, therefore, twofold aim:
to find energy distribution of a quantum particle as a function of ``coordinates in
space-time" and to show how it emerges due to the procedure of the objective quantum measurement.

Generally, it is important to understand that the problem of identification of
physical objects is the root problem even in classical physics and that its recognition
gave to Einstein the key to formalization of the relativistic kinematics and dynamics.
Indeed, only assuming the possibility to detect locally an approximate
coincidence of two pointwise events of a different nature it is
possible to build full kinematic scheme and the physical geometry of
space-time \cite{Einstein1,Einstein2}. As such the ``state" of the
local clock gives us local coordinates - the ``state" of the
incoming train. In the classical case the notions of the ``clock"
and the ``train" are intuitively clear and approximately may be identified with
material points or even with space-time points. This supports the illusion that
material bodies present in space-time (Einstein emphasized that it is not so!).
Furthermore, Einstein especially notes that he did not discuss the inaccuracy of the
simultaneity of two {\it approximately coinciding events} that should
be overcame by some abstraction \cite{Einstein1}. This abstraction
is of course the neglect of finite sizes (and all internal degrees
of freedom) of the both real clock and train. It gives the
representation of these ``states" by mathematical points in
space-time. Thereby the local identification of two events is the
formal source of the classical relativistic theory. The world line is the second
essential component of the relativistic kinematics and dynamics. Namely, we hope
(and our belief is based on macroscopic experience) that the evolution of material
point obeys some dynamical law and even without intermediate measurements of
coordinates we know its space-time position (determinism). The world line is a
mathematical expression of the ``fate" of some material point and two points have
two different world lines. In that sense we have identification of each material
point with mathematical points of the world line.

However quantum object requires especial embedding in space-time and its the
identification with space-time point is impossible since the localization of
quantum particles is state-dependent. Hence the identification of quantum objects
requires a physically motivated operational procedure with corresponding
mathematical description. The quantum problem of localization in space-time
is rooted in the linearity of the fundamental wave equations. Solution of such
equations can not keep stable coherent superposition. This leads to impossibility
to define the trajectory of quantum particles and their identification. Nevertheless,
trajectories are clearly seen on the photo plates, in Wilson, or in bubble chamber.
What the commonly used expression ``we see" really means? What we see and how?
We see the result of billion interactions of a moving particle with ionizing atoms.
The droplets of mist or bubbles are condensing in the vicinity of ionized atoms
and they (droplets or bubbles) shape the trace of the particle in the medium that
we identify as a ``trajectory". Subjectively, the quantum measurement is a human
observation (identification plus comparison)
and a numerical encoding of the result of the comparison, i.e. in fact the reply on
quantum question: ``yes" or ``no". The chain of such quantum questions takes the place
in the example given above. What is the objective content of the quantum measurement, if any?

The objective quantum measurement is in fact invariant
quantum geometry without any mention of human presence. Two notions
comprising the basis of the general measurement scheme have been used:
velocity of the GCS deformation and its
qubit encoding. The quantum measurement problem should be invariantly
formulated as a comparison of quantum LDV's inherently connected with GCS (instead of
the comparison of quantum states themselves) with help of their affine parallel transport
in $CP(N-1)$ and encoding the result of this comparison by the qubit spinor whose
components in infinitesimally close points of the state space define the dynamical
space-time \cite{Le2,Le3}. In order to reach objective character of the whole procedure
the qubit spinor components should be given by the ``invariant state reduction
procedure" too. It is possible to say that objective character of the quantum measurement
is inherently related to the space-time structure that replaces the notion of ``observer"
in the framework of super-relativity. This scheme requires an essential reconstruction
of all formal quantum apparatus.
\section{Super-relativity}
The concept of super-relativity \cite{Le1} arose as a development of the
Fock's idea of ``relativity to measuring device". This idea may be
treated originally as generalization of the relativity concept in space-time to
the some ``functional relativity" in the state space. However the power
of this program is very limited in comparison with power of Einstein's concepts of special or general relativity. The main reason is that the notion of the ``measuring device" could not be correctly formulated in the own framework of the standard quantum theory. Some additional and, in fact, outlandish classical ingredients should be involved. It
definitely related to well known ``measurement problem" in quantum theory.

In order to avoid these difficulties, the pure quantum construction
of dynamics in the projective Hilbert space has been proposed instead of ``realistic"
usage of ``classical analogy method". This choice is dictated by well known
properties of pure quantum states but now the role of these states should be
quite different in comparison with original Schr\"odinger scheme. I use
these states initially in pure abstract manner in order to develop unitary
classification of quantum motions: this leads to the important notions of the
local dynamical variables (LDV's) represented by tangent vector fields to $CP(N-1)$.
Two kinds of these fields (generators of coset and isotropy sub-group transformations) are similar to Goldstone's and Higgs fields, therefore there is some possibility to
identify the geometry of the unitary group and quantum forces. Thus, in clear reason, arose an ambitious program that has been called ``super-relativity". Definitely, the abstract classification
of unitary motions is not sufficient for physical theory since it should be connected
with measuring in space-time.

Ordinary approach of the relativistic QFT assumes that Poincar\'e  group should be
linearly represented in some Hilbert space by dynamical variables (linear operators)
acting in this Hilbert space. This approach leads to a lot of the conceptual and
technical problems \cite{Schroer}. Technically, super-relativity \cite{Le1,Le2,Le3,Le4}
uses an ``inverse representation", namely: unitary dynamical group $SU(N)$ should be non-linearly represented by non-local soliton-like field object associated with quantum particle.

In other words, I assume that ``initial" state of quantum system may be represented by vectors the Hilbert space of action states (AS). The action states are used here as
states of hidden stationary ``elementary" quantum motions without any
reference to its placement or momentum (background independent formulation).
The rays $\{|F>\}$ of AS serve merely for ``functional" localization of quantum system.
The classification of unitary motions of the rays is based on pure geometric structure
of the $SU(N)$, its isotropy sub-group  $H[|F>]=U(1) \times U(N-1)$ of some AS
vector $|F>$, and the coset structure $G/H[|F>]=SU(N)/S[U(1) \times U(N-1)]=CP(N-1)$
of the unitary transformations taking the place of ``quantum force" \cite{Le1,Le2,Le3,Le4}.

The dynamical space-time (DST) arises as the section of the tangent fiber bundle
over $CP(N-1)$ and it has a ``granular structure that respects Lorentz symmetry"
only locally \cite{Bonder}.  This DST is realized as coordinates $x^{\mu}$ manifold
for energy distribution due to single process of quantum evolution in two
infinitesimally close points of trajectory in $CP(N-1)$ (say, due to a measurement
of some local dynamical variable). I used the affine parallel transport of local
Hamiltonian $CP(N-1)$ along this trajectory in order to be sure that at two different
states one has the ``same" quantum system (self-conservation or self-identification
as the reference to infinitesimally close previous state in the Cartan's sense).
Quantum measurement is encoded by the qubit spinor (formal two-level system
with eigen-states $|yes>, |no>$ of the quantum question). One of them will be associated
with tangent vector to $CP(N-1)$ shows the direction and the speed of evolution
from one general coherent state (GCS) to another under the coset transformations,
the second one will be associated with the normal vector to $CP(N-1)$ representing the
transformations of isotropy group of some GCS.

It is well known that two infinitesimally close spinors may be formally connected by
infinitesimal Lorentz transformation. I assumed that this relation may have not merely
mathematical sense, but, being applied to the qubit spinor encoding result of
quantum interaction (self-interaction) used for a ``measurement", as the real reason
of the four-dimension nature of dynamical space-time manifold.

\section{The universality of ``corpuscule-wave duality" and second quantization}
The statistical analysis of the energy distribution is the base of the black body
radiation \cite{Planck1} and the Einstein's theory of the light emission and absorption
\cite{Einstein_Q}. This conceptual line was logically finished by Dirac in his
method of the second quantization \cite{Dirac1}. This approach is perfectly fits to
many-body weakly interacting quantum systems and it was assumed that the ``corpuscule-wave duality" is universal. However the application of this method to single quantum ``elementary" particles destroys this harmony. Physically it is clear why: quantum particle is self-interacting system and this interaction is at least of the order of its
rest mass. Since the nature of the mass is the open problem we do not know
the energy distribution in quantum particles up to now. Notice, Einstein \cite{EPR} and
Schr\"odinger \cite{Schr} treated the statistical fundament of quantum theory as a
perishable and temporal. A long time it is was assumed that the dynamical model may
be found in the framework of the string theory, but the epitaph to string
theory \cite{Schroer} only subscribes the deep crisis in particle physics as whole.
One of the aim of this article to show how it is possible to find energy distribution
in single non-local quantum particle - ``lump".

It is remarkable that Blochintzev more then 50 years before discussed the universality
of wave - particle ``duality" connected with the method of second quantization \cite{Bl1,Bl2}. It was shown that such universality is generally broken for interacting quantum systems. Namely, attempt to represent two interacting boson fields as the set of free quantum oscillators leads to two types of oscillators: quantized and non-quantized. The second one arises under simple relation $g > \frac{mMc^2}{h^2}$ between coupling constant $g$ and masses $m$ and $M$ of two scalar fields. For such intensity of coupling we obtain a field state without any counterparting ``particles". For self-interacting scalar field of mass $m$ the intensity of self-interaction $g$ leads to breakdown of the universality of the wave - particle ``duality" if it is larger than the inverse square of the Compton wavelength: $g  > \frac{m^2c^2}{h^2}=\frac{1}{\lambda^2_C}$.
In order to build dynamical in lieu of statistical quantum theory let me assume
that there exist some \textbf{ideal} ``elementary" quantum states of internal motion
with quantized finite action without any connection to environment, i.e. without
interaction (background independent quantum states)  \cite{Le2}. Namely, the action
(not energy) of some ``elementary" quantum motion is assumed to be primary quantized
but energy distribution should be established during dynamics in state space. The
states of these ``elementary" quantum motion (action states) play the role similar
to the role of inertial frames in classical physics.

POSTULATE 1.

\noindent {\it There are elementary quantum states $|\hbar a>,
a=0,1,...$ belonging to the Fock space of an abstract Planck
oscillator whose states correspond to the quantum motions with given
number of Planck action quanta}.

 One may image some {\it ``elementary quantum states"
(EAS) $|\hbar a>$ as a quantum motions with entire number $a$ of the
action quanta}. These $a,b,c,...= 0, 1, 2,...$ take the place of the ``principle
quantum number" serving as discrete indices $0 \leq a,b,c... <~
\infty$.

Each action state of quantum system is a quantum motion in some ``dynamical order"  defined, say, be some Lagrangian or action functional.
The AS vector of this quantum motion may be represented by the generalized coherent state (GCS)
\begin{eqnarray}\label{1}
|F>=\sum_{a=0}^{\infty} f^a| \hbar a>,
\end{eqnarray}
where $|\hbar a>=(a!)^{-1/2} ({\hat \eta^+})^a|\hbar 0>$
and, thereby, may be treated as ``order parameter" belonging to Hilbert space $\cal{H}$ - ``the space of the order parameter".

These quantum AS of motion do not gravitate since they are ``pre-matter" and don't posses such fundamental physical attributes like position, momentum and mass/energy by itself. Therefore their linear superposition is robust and the rays of these GCS will be the main building blocks of the model. Probably quark's multiplete is one of the such kind of non-observable quantum states. Only velocities of variation of these states given by local dynamical variables correspond to ``materialized" quantum states. In order to analyze the quantum dynamics of such states we should use unitary kinematics for classification of the GCS motions \cite{Le1}. Such excitations create quantum excitations like particles, solitons, unparticles, etc., under some conditions that should be especially established and studied.

The introduction of the cosmic potential $\Phi_U = c^2$ is the simplest way to  ``materialization" of quantum motions. It forms some global vacuum $|\Phi_U> = |\hbar 0>$ whose perturbation by the Hamiltonian
\begin{eqnarray}\label{2}
\hat{H}=\hbar \omega {\hat \eta^+} {\hat \eta}=mc^2 {\hat \eta^+} {\hat \eta}=\omega \hat{S_P}= \frac{mc^2}{\hbar}\hat{S_P},
\end{eqnarray}
where the Planck's action quanta
operator $\hat{S_P}=\hbar {\hat \eta^+} {\hat \eta}$ with the
spectrum $S_a=\hbar a$ in the separable Hilbert space $\cal{H}$ is merely the simplest case of more general expression of the action operator
\begin{eqnarray}\label{3}
\hat{S}=\hbar A({\hat \eta^+} {\hat \eta}).
\end{eqnarray}
where $A$ is some analytic function of ordinary Bose operators of creation-annihilation $\hat \eta^+$ and $\hat \eta$.

Formally these oscillations may be represented by the superposition
in infinite dimension manifold of the \emph{Planck's oscillators of
action}. I discuss here $N$-level model of finite quantum action in some system whose states $|F>$ correspond to extremals of some least action problem and describe stationary quantum motion. This relative (local) vacuum
of some problem is not necessarily the state with minimal energy, it
is a state with an extremal of some action functional.
Thus $G=SU(N)$ is dynamical group of the order parameter and its geometry is the base for the classification of GCS motions \cite{Le1}.
\section{Geometry of the quantum evolution and/or measurement}
Quantum evolution has generally unitary and/or non-unitary character. The transition from the unitary regime to the non-unitary one is used frequently for the connection of micro- to macrophysics in relation with the measurement problem in quantum theory; particulary with the definition of so-called ``quantum measurement machine" \cite{Wezel}. I, however, think that we should recognize after all the rights of quantum system to have the objective sense without any reference to necessity of ``observer" or ``quantum measurement machine".
I propose the natural geometric mechanism of the unitary breakdown, it is in fact the reconstruction of the unitary symmetry under non-linear dynamical mapping \cite{TFD}.
This approach leads to following result: \textbf{objective character of the quantum theory related to self-consistent introduction of the dynamical space-time.}
The internal unitary classification of quantum motions has been already proposed \cite{Le1}.
Physical motivation given 12 years before may be reinforced now by appeal to
dynamical effects of Lorentz transformations as follows.

One may assume that approximately the following equation for ``free" electrons is correct
\begin{eqnarray}\label{4}
|electron, x>=|electron, y> =
|F(x)>=|F(y)> or \cr |F(x)>=|F(L x + a)> =|U_{H[|F(x)>]} |F(x)>,
\end{eqnarray}
where $U_{H[|F(x)>]}$ unitary transformations from the isotropy subgroup
of $H[|F(x)>]=U(1) \times U(N-1)$. Since realization of the this subgroup is
state-dependent the embedding (parametrization) of subgroup $H[F(x)]$ and the coset
transformations $G/H[|F(x)>]=SU(N)/S[U(1) \times U(N-1)] = CP(N-1)[|F(x)>]$ is state-dependent too \cite{Le3}.

It has been shown that dynamical reconstruction of the unitary symmetry $SU(N)$ leads to creation of non-local ``lump" of surrounding field $\Omega^{\alpha}(x)$ belong to adjoint representation of $SU(N)$. Physically it is motivated by the fact that states of two electrons may be identical only asymptotically since they interact at finite distance and mutually perturb their quantum states. Moreover, quantum electron is self-interacting and therefore even single electron in point  $x$ and in point $y$ is not generally ``same". Strictly speaking it means that naive quantization of interacting (and self-interacting) systems leads if not to contradictions, but at least to new ``unparticle physics" predicted by Blochintzev more then 50 years before \cite{Bl1,Bl2}. Formally it means that the quantum state $|F(L x +a )> $ of electron in the point $y=Lx+a$
could not be obtained by action of the isotropy group of $|F(x)>$ representing Poincar\'e transformation. Generally one has rather
\begin{eqnarray}\label{5}
|F(x)> \rightarrow |F(y)> = |F(L x +a )> =(G/H) |F(x)> \cr \neq U_{H[|F(x)>]} |F(x)>.
\end{eqnarray}
However the last equation could not be globally exact since the coset parametrization is state dependent. Only locally in $CP(N-1)$ and in the space-time it is may be correct.
It is the consequence of the fact that an actual (not mental) quantum motion from the one space-time point to another leads to dynamical effects. In order to take into account such effect I assumed that two postulates should be used:

1. ``Super-relativity" : \textbf{coset deformation of quantum state may be created and compensated by some physical fields} and

 2. ``Dynamical space-time structure emergence" :\textbf{quantum measurement of the local dynamical variables may be encoded by the local Lorentz transformations of the qubit spinor representing the state of two-level detector into local dynamical space-time}. Thereby local space-time structure accompanying quantum dynamics may be established.

Therefore the differential form of the equivalence
\begin{eqnarray}\label{6}
|F(x+dL x)> = |F(x)> + d(G/H) |F(x)>
\end{eqnarray}
should be used. I formulate the equivalence problem (\ref{6}) with help field equations for the $SU(N)$ parameters $\Omega^{\alpha}, (1 \leq \alpha \leq N^2 - 1)$
providing the affine parallel transport of the Hamiltonian field
$H^i=\hbar \Omega^{\alpha}\Phi^i_{\alpha}$ \cite{Le2,Le4,Le5}.

Furthermore, since the space-time coordinates do not have physical meaning by itself, they should be introduced in self-consistent manner, i.e, dynamical reconstruction of the global unitary symmetry which is broken due to non-linear dynamical mapping lead to infinitesimal transformations of the surrounding ``fields shell" $\Omega(x)$ if one expresses the conservation law of the local Hamiltonian in the form of the affine parallel transport in $CP(N-1)$. Then the quantum measurement of the LDV being encoded with help infinitesimal Lorentz transformations of qubit spinor leads to the emergence of the dynamical space-time. Non-local soliton-like objects arising due to affine parallel transport of the LDV's breaks
global Lorentz symmetry but it may be restored locally in dynamical space-time.

Let me assume that ``ground
state" $|G>=\sum_{a=0}^{N-1} g^a|\hbar a>$ is a solution of some the least action problem.
Since any action state $|G>$ has (in appropriate basis) the isotropy group
$H=U(1)\times U(N)$, only the coset transformations $G/H=SU(N)/S[U(1)
\times U(N-1)]=CP(N-1)$ effectively act in $\cal{H}$. Therefore the
ray representation of $SU(N)$ in $C^N$, in particular, the embedding
of $H$ and $G/H$ in $G$, is a state-dependent parametrization.
As I wrote before, we should use the local from of the equivalence principle (\ref{5}).
Technically the local $SU(N)$ unitary classification of the quantum motions requires the transition from the matrices of Pauli $\hat{\sigma}_{\alpha},(\alpha=1,...,3)$, Gell-Mann $\hat{\lambda}_{\alpha},(\alpha=1,...,8)$, and in general $N \times N$ matrices $\hat{\Lambda}_{\alpha}(N),(\alpha=1,...,N^2-1)$ of $AlgSU(N)$ to the tangent vector fields to $CP(N-1)$ in local coordinates \cite{Le1}.
The transition to the local coordinates is in fact non-linear \textbf{dynamical mapping}
onto $CP(N-1)$ \cite{TFD}.
Hence, there is a diffeomorphism between the space of the rays
marked by the local coordinates in the map
 $U_j:\{|G>,|g^j| \neq 0 \}, j>0$
\begin{equation}\label{7}
\pi^i_{(j)}=\cases{\frac{g^i}{g^j},&if $ 1 \leq i < j$ \cr
\frac{g^{i+1}}{g^j}&if $j \leq i < N-1$}
\end{equation}
and the group manifold of the coset transformations
$G/H=SU(N)/S[U(1) \times U(N-1)]=CP(N-1)$ and the isotropy group of the corresponding ray with local coordinates (\ref{6}).
This diffeomorphism is provided by the coefficient functions $\Phi^i_{\alpha}$
\begin{equation}\label{8}
\Phi_{\sigma}^i = \lim_{\epsilon \to 0} \epsilon^{-1}
\biggl\{\frac{[\exp(i\epsilon \hat{\lambda}_{\sigma})]_m^i g^m}{[\exp(i
\epsilon \hat{\lambda}_{\sigma})]_m^j g^m }-\frac{g^i}{g^j} \biggr\}=
\lim_{\epsilon \to 0} \epsilon^{-1} \{ \pi^i(\epsilon
\hat{\lambda}_{\sigma}) -\pi^i \}
\end{equation}
of the local generators
\begin{equation}\label{9}
D_{\sigma}=\Phi_{\sigma}^i \frac{\partial}{\partial \pi^i} + c.c.
\end{equation}
comprise of non-holonomic basis of $CP(N-1)$ \cite{Le1}.
In fact the definition (\ref{8}) is equivalent to the first variation of the dynamical mapping (see (7.4.4) in \cite{TFD}). This provides the local projection of the unitary group $SU(N)$ onto the base manifold $CP(N-1)$.
\subsection{New definition of the state vector}
The general spirit of the newly defined quantum evolution requires a new construction of the state vector $|\Psi>$. It will be associated now with a velocity of the GCS variation and,
thereby, with vector field of differential operators in $\pi^i$ with coefficient functions given by $\Phi^i_{\alpha}$. In that sense this construction is similar to the second quantization scheme where state vector has an operator nature too.

The coordinates of the ``ground
state" $|G>=\sum_{a=0}^{N-1} g^a|\hbar a>$ may be expressed in local coordinates
as follows: for $a=0$ one has
\begin{eqnarray}\label{10}
g^0(\pi^1_{j(p)},...,\pi^{N-1}_{j(p)})=(1+
\sum_{s=1}^{N-1}|\pi^s_{j(p)}|^2)^{-1/2}
\end{eqnarray}
and for $a: 1\leq a = i \leq N-1$ one has
\begin{eqnarray}\label{11}
g^i(\pi^1_{j(p)},...,\pi^{N-1}_{j(p)})= \pi^i_{j(p)}(1+
\sum_{s=1}^{N-1}|\pi^s_{j(p)}|^2)^{-1/2}.
\end{eqnarray}
Then the velocity of the ground state evolution relative the length parameter in $CP(N-1)$ playing the role of the ``world
time" $\tau$ is given by the formula
\begin{eqnarray}\label{12}
|\Psi> \equiv |T> =\frac{d|G>}{d\tau}=\frac{\partial g^a}{\partial
\pi^i}\frac{d\pi^i}{d\tau}|\hbar a>+\frac{\partial g^a}{\partial
\pi^{*i}}\frac{d\pi^{*i}}{d\tau}|\hbar a> \cr
=|T_i>\frac{d\pi^i}{d\tau}+|T_{*i}>\frac{d\pi^{*i}}{d\tau}=H^i|T_i>+H^{*i}|T_{*i}>,
\end{eqnarray}
is the tangent vector to the evolution curve $\pi^i=\pi^i(\tau)$,
where
\begin{eqnarray}\label{13}
|T_i> = \frac{\partial g^a}{\partial \pi^i}|\hbar a>=T^a_i|\hbar a>,
\quad |T_{*i}> = \frac{\partial g^a}{\partial
\pi^{*i}}|\hbar a>=T^a_{*i}|\hbar a>.
\end{eqnarray}
Thereby state vector $|\Psi> \equiv |T>$ giving velocity of evolution,
is represented by the tangent vector to the projective Hilbert space $CP(N-1)$ in local  coordinates
$\pi^k_{(j)}=\frac{g^k}{g^j}$ of the quantum states. The parallel
transport of $|\Psi>$ is required to be in agreement with the Fubini-Study metric
\begin{equation}\label{14}
G_{ik^*} = [(1+ \sum |\pi^s|^2) \delta_{ik}- \pi^{i^*} \pi^k](1+
\sum |\pi^s|^2)^{-2} \label{FS}.
\end{equation}
Then the affine connection
\begin{eqnarray}\label{15}
\Gamma^i_{mn} = \frac{1}{2}G^{ip^*} (\frac{\partial
G_{mp^*}}{\partial \pi^n} + \frac{\partial G_{p^*n}}{\partial
\pi^m}) = -  \frac{\delta^i_m \pi^{n^*} + \delta^i_n \pi^{m^*}}{1+
\sum |\pi^s|^2} \label{Gamma}
\end{eqnarray}
takes the place of the gauge potential of the non-Abelian type
playing the role of the covariant instant renormalization of the
dynamical variables during general transformations of the quantum
self-reference frame \cite{Le2}.

Velocity of the $|\Psi>$ variation is given by the equation
\begin{eqnarray}\label{16}
|A> &=&\frac{d|\Psi>}{d\tau} \cr &=&
(B_{ik}H^i\frac{d\pi^k}{d\tau}+B_{ik^*}H^i\frac{d\pi^{k*}}{d\tau}
+B_{i^*k}H^{i^*}\frac{d\pi^k}{d\tau} +B_{i^*
k^*}H^{i^*}\frac{d\pi^{k*}}{d\tau})|N>\cr &+&
(\frac{dH^s}{d\tau}+\Gamma_{ik}^s
H^i\frac{d\pi^k}{d\tau})|T_s>+(\frac{dH^{s*}}{d\tau}+\Gamma_{i^*k^*}^{s*}
H^{i*}\frac{d\pi^{k*}}{d\tau})|T_{s*}>,
\end{eqnarray}
where I introduce the matrix $\tilde{B}$ of the second quadratic
form whose components are defined by following equations
\begin{eqnarray}\label{17}
B_{ik}|N> =\frac{\partial |T_i>}{\partial \pi^k}-\Gamma_{ik}^s|T_s>,
\quad B_{ik^*}|N> = \frac{\partial |T_i>}{\partial \pi^{k*}} \cr
B_{i^*k}|N> =\frac{\partial |T_{i*}>}{\partial \pi^k}, \quad B_{i^*
k^*}|N> = \frac{\partial |T_{i*}>}{\partial
\pi^{k*}}-\Gamma_{i^*k^*}^{s*}|T_{s*}>
\end{eqnarray}
through the state $|N>$ normal to the ``hypersurface'' of the ground
states. Assuming that the ``acceleration'' $|A>$ is gotten by the
action of some linear Hamiltonian $\hat{H}_S$ describing the
evolution (say, during a measurement), one has the ``Schr\"odinger equation
of evolution"
\begin{eqnarray}\label{18}
\frac{d|\Psi>}{d\tau}&=&-i\hat{H}_S|\Psi> \cr
&=&(B_{ik}H^i\frac{d\pi^k}{d\tau}+B_{ik^*}H^i\frac{d\pi^{k*}}{d\tau}
+B_{i^*k}H^{i^*}\frac{d\pi^k}{d\tau} +B_{i^*
k^*}H^{i^*}\frac{d\pi^{k*}}{d\tau})|N> \cr &+&
(\frac{dH^s}{d\tau}+\Gamma_{ik}^s
H^i\frac{d\pi^k}{d\tau})|T_s>+(\frac{dH^{s*}}{d\tau}+\Gamma_{i^*k^*}^{s*}
H^{i*}\frac{d\pi^{k*}}{d\tau})|T_{s*}>.
\end{eqnarray}
I should emphasize that the ``world time" here is non identical to the ``world time"
of Stueckelberg-Horwitz since it is the time of evolution from one generalized
coherent state (GCS) to another. In fact it is proportional to the length of the
trajectory in $CP(N-1)$ in the sense of Fubini-Study metric. Probably it is better
to call it ``omnipresent time" in the sense that under identical initial conditions
the rate of the quantum evolution at any place of Universe is identical.

Thereby the unitary evolution of the action amplitudes generated by
\begin{eqnarray}\label{19}
\hat{U}(\tau)=e^{i \tau \Omega^{\alpha}\hat{\lambda}_{\alpha}}=e^{i \tau \hat{H}}
\end{eqnarray}
leads in general to the non-unitary evolution of the tangent
vector to $CP(N-1)$ associated with ``state vector" $|\Psi>$ since
the Hamiltonian $\hat{H}_S$ is non-Hermitian and its expectation
values are as follows:
\begin{eqnarray}\label{20}
<N|\hat{H}_S|\Psi>&=&
i(B_{ik}H^i\frac{d\pi^k}{d\tau}+B_{ik^*}H^i\frac{d\pi^{k*}}{d\tau}
+B_{i^*k}H^{i^*}\frac{d\pi^k}{d\tau} +B_{i^*
k^*}H^{i^*}\frac{d\pi^{k*}}{d\tau}),\cr <\Psi|\hat{H}_S|\Psi>&=&
iG_{p^*s}(\frac{dH^s}{d\tau}+\Gamma_{ik}^s
H^i\frac{d\pi^k}{d\tau})H^{p*}+iG_{ps^*}(\frac{dH^{s*}}{d\tau}+\Gamma_{i^*
k^*}^{s*} H^{i^*}\frac{d\pi^{k*}}{d\tau})H^p\cr
&=&i<\Psi|\frac{d}{d\tau}|\Psi>.
\end{eqnarray}
The minimization of the $|A>$ under the transition from point $\tau$
to $\tau+d\tau$ may be achieved by the annihilation of the
tangent component
\begin{equation}\label{21}
\frac{dH^s}{d\tau}+\Gamma_{ik}^s H^i\frac{d\pi^k}{d\tau}=0, \quad
\frac{dH^{s*}}{d\tau}+\Gamma_{i^* k^*}^{s*}
H^{i^*}\frac{d\pi^{k*}}{d\tau}=0
\end{equation}
i.e. under the condition of the affine parallel transport of the
Hamiltonian vector field. The last equations in (\ref{20}) shows that the
affine parallel transport of $H^i$ agrees with Fubini-Study metric
(\ref{14})  leads to Berry's ``parallel transport" of $|\Psi>$.
\section{Dynamical space-time instead of ``observer"}
 Functionally invariant construction should be used instead of ``observer". I have assumed that the quantum measurement of the LDV being encoded with help infinitesimal Lorentz transformations of qubit spinor leads to emergence of the dynamical space-time that takes the place of the objective ``quantum measurement machine" formalizing the process of numerical encoding the results of comparisons of LDV's. Two these procedures are described below.

\subsection{LDV's comparison}
Local coordinates $\pi^i$ of the GCS in $CP(N-1)$ give reliable geometric tool for
the description of quantum dynamics during interaction or self-interaction. This
leads to evolution of GCS and that may be used in measuring process. Two essential components of any measurement are
identification and comparison. The Cartan's idea of ``self-identification" by the reference to the previous infinitesimally close GCS has been used. Thereby, LDV is now a new essential element of quantum dynamics. We should be able to compare some LDV at two infinitesimally
close GCS represented by points of $CP(N-1)$. Since LDV's are vector fields on $CP(N-1)$,
the most natural mean of comparison of the LDV's is affine parallel transport
agrees with Fubini-Study metric \cite{Le1}.

This parallel transport being applied to the Hamiltonian vector field
$H^i=\hbar \Omega^{\alpha}\Phi^i_{\alpha}$
\begin{eqnarray}\label{22}
\frac{\delta H^k}{\delta \tau} &= &\hbar \frac{\delta
(\Phi^k_{\alpha} \Omega^{\alpha})}{\delta \tau}=0,
\end{eqnarray}
leads to the equation for differentials of the field parameters $\Omega^{\alpha}$ of the $SU(N)$ that may be expressed as follows:
\begin{eqnarray}\label{23}
\delta \Omega^{\alpha}=-
(\Gamma^m_{mn} \Phi_{\beta}^n+\frac{\partial
\Phi_{\beta}^n}{\partial \pi^n}) \Omega^{\alpha}\Omega^{\beta} \delta \tau.
\end{eqnarray}

\subsection{Encoding the results of comparison}
The results of the comparison of LDV's should be formalized by numerical encoding.
Thus one may say that ``LDV has been measured". The invariant encoding is based on
the geometry of $CP(N-1)$ and LDV dynamics, say, dynamics of the local Hamiltonian field.
Its affine parallel transport expresses the self-conservation of quantum lump associated with ``particle". In order to build the qubit spinor $\eta$ of the quantum question
$\hat{Q}$ \cite{Le5} two orthogonal vectors $\{|N>,|\Psi>\}$ have been used.
I will use following equations
\begin{eqnarray}\label{24}
\eta=\left(
  \begin{array}{cc}
    \alpha_{(\pi^1,...,\pi^{N-1})}  \\
    \beta_{(\pi^1,...,\pi^{N-1})} \\
  \end{array}
\right) = \left(
  \begin{array}{cc}
    \frac{<N|\hat{H}|\Psi>}{<N|N>}  \\
    \frac{<\Psi|\hat{H}|\Psi>}{<\Psi|\Psi>} \\
  \end{array}
\right)
\end{eqnarray}
for the
measurement of the Hamiltonian $\hat{H}$ at corresponding GCS.
Then from the infinitesimally close GCS
$(\pi^1+\delta^1,...,\pi^{N-1}+\delta^{N-1})$, whose shift is
induced by the interaction used for a measurement, one get a close
spinor $\eta+\delta \eta$ with the components
\begin{eqnarray}\label{25}
\eta + \delta \eta =\left(
  \begin{array}{cc}
    \alpha_{(\pi^1+\delta^1,...,\pi^{N-1}+\delta^{N-1})}  \\
    \beta_{(\pi^1+\delta^1,...,\pi^{N-1}+\delta^{N-1})} \\
  \end{array}
\right) = \left(
  \begin{array}{cc}
    \frac{<N|\hat{H'}|\Psi>}{<N|N>}  \\
    \frac{<\Psi|\hat{H'}|\Psi>}{<\Psi|\Psi>}
  \end{array}
\right).
\end{eqnarray}
Here $\hat{H}=\hbar \Omega^{\alpha}\hat{\lambda}_{\alpha}$ is the lift of Hamiltonian tangent vector field $H^i=\hbar \Omega^{\alpha} \Phi^i_{\alpha}$ from $(\pi^1,...,\pi^{N-1})$ and $\hat{H'}=\hbar(\Omega^{\alpha}+\delta \Omega^{\alpha}) \hat{\lambda}_{\alpha}$ is the lift of the same tangent vector field parallel transported from the infinitesimally close point
$(\pi^1+\delta^1,...,\pi^{N-1}+\delta^{N-1})$ back to the
$(\pi^1,...,\pi^{N-1})$ into the adjoint representation space. Then one finds
\begin{eqnarray}\label{26}
 \delta \eta =\left(
  \begin{array}{cc}
    \alpha_{(\pi^1+\delta^1,...,\pi^{N-1}+\delta^{N-1})}-\alpha_{(\pi^1,...,\pi^{N-1})}  \\
    \beta_{(\pi^1+\delta^1,...,\pi^{N-1}+\delta^{N-1})}-\beta_{(\pi^1,...,\pi^{N-1})} \\
  \end{array}
\right) = \left(
  \begin{array}{cc}
    \frac{<N|\delta \Omega^{\alpha} \hat{\lambda}_{\alpha}|\Psi>}{<N|N>}  \\
    \frac{<\Psi|\delta \Omega^{\alpha} \hat{\lambda}_{\alpha}|\Psi>}{<\Psi|\Psi>} \\
  \end{array}
\right),
\end{eqnarray}
where one should find how the affine parallel transport connected
with the variation of coefficients $\Omega^{\alpha}$ in the
dynamical space-time associated with quantum question $\hat{Q}$.
The covariance relative transition from one GCS to another
\begin{eqnarray}\label{27}
(\pi^1_{j(p)},...,\pi^{N-1}_{j(p)}) \rightarrow
(\pi^1_{j'(q)},...,\pi^{N-1}_{j'(q)})
\end{eqnarray}
and the covariant differentiation (relative Fubini-Study metric) of
vector fields provides the objective character of the ``quantum
question" $\hat{Q}$ and, hence, the quantum measurement. This serves
as a base for the construction of the dynamical space-time as it
will be shown below.

Each quantum measurement consists of the procedure of encoding
of quantum dynamical variable into state of a ``pointer" of ``macroscopic measurement machine" \cite{Wezel}. Quantum lump takes the place of such extended ``pointer".
This extended pointer may be mapped onto dynamical space-time if one assumes
that transition from one GCS to another is accompanied by dynamical
transition from one Lorentz frame to another attached to adjacent point of the ``pointer".
Thereby, infinitesimal Lorentz transformations define small
``dynamical space-time'' coordinates variations. It is convenient to take
Lorentz transformations in the following form
\begin{eqnarray}\label{28}
ct'=ct+(\vec{x} \vec{a}) \delta \tau \cr
\vec{x'}=\vec{x}+ct\vec{a} \delta \tau
+(\vec{\omega} \times \vec{x}) \delta \tau
\end{eqnarray}
where I put
$\vec{a}=(a_1/c,a_2/c,a_3/c), \quad
\vec{\omega}=(\omega_1,\omega_2,\omega_3)$ \cite{G} in order to have
for $\tau$ the physical dimension of time. The expression for the
``4-velocity" $ V^{\mu}$ is as follows
\begin{equation}\label{29}
V^{\mu}=\frac{\delta x^{\mu}}{\delta \tau} = (\vec{x} \vec{a},
ct\vec{a}  +\vec{\omega} \times \vec{x}) .
\end{equation}
The coordinates $x^\mu$ of points in dynamical space-time serve here merely for the parametrization of the energy distribution in the ``field
shell'' arising under its motion according to non-linear field
equations \cite{Le1,Le3}. It is interesting to note that ``4-velocity" $ V^{\mu}$
and ``4-acceleration"
\begin{equation}\label{30}
A^{\mu}=\frac{\delta^2 x^{\mu}}{\delta \tau^2} = (\vec{a}[ct\vec{a}+\vec{\omega} \times \vec{x}],
\vec{a}(\vec{a}\vec{x})  +\vec{\omega}[ct\vec{a}+ \vec{\omega} \times \vec{x}]) .
\end{equation}
have zero value at the origin and increase in all space-time directions.
Probably, it somehow connected with observable expansion of Universe, but this topic
is outside of our envision.

Any two infinitesimally close spinors $\eta$ and $\eta+\delta
\eta$ may be formally connected with infinitesimal ``Lorentz spin transformations
matrix'' \cite{G}
\begin{eqnarray}\label{31}
L=\left( \begin {array}{cc} 1-\frac{i}{2}\delta \tau ( \omega_3+ia_3 )
&-\frac{i}{2}\delta \tau ( \omega_1+ia_1 -i ( \omega_2+ia_2)) \cr
-\frac{i}{2}\delta \tau
 ( \omega_1+ia_1+i ( \omega_2+ia_2))
 &1-\frac{i}{2}\delta \tau( -\omega_3-ia_3)
\end {array} \right).
\end{eqnarray}
I have assumed that there is not only formal but dynamical reason for such transition
when Lorentz reference frame moves together with GCS.
Then ``quantum accelerations" $a_1,a_2,a_3$ and ``quantum angle velocities" $\omega_1,
\omega_2, \omega_3$ may be found in the linear approximation from
the equation
\begin{eqnarray}\label{32}
\eta+\delta \eta = L \eta
\end{eqnarray}
as functions of the qubit spinor components of the quantum question
depending on local coordinates $(\pi^1,...,\pi^{N-1})$ involved in the $\delta \Omega^{\alpha}$ throughout field equations (\ref{40})
\begin{eqnarray}\label{33}
 \delta \eta = L \eta - \eta \cr
  = \left( \begin {array}{cc} -\frac{i}{2}\delta \tau ( \omega_3+ia_3 )
&-\frac{i}{2}\delta \tau ( \omega_1+ia_1 -i ( \omega_2+ia_2)) \cr
-\frac{i}{2}\delta \tau
 ( \omega_1+ia_1+i ( \omega_2+ia_2))
 &-\frac{i}{2}\delta \tau( -\omega_3-ia_3)
\end {array} \right) \cr \left(
  \begin{array}{cc}
    \frac{<N|\Omega^{\alpha} \hat{\lambda}_{\alpha}|\Psi>}{<N|N>}  \\
    \frac{<\Psi|\Omega^{\alpha} \hat{\lambda}_{\alpha}|\Psi>}{<\Psi|\Psi>}
  \end{array}
\right).
\end{eqnarray}
Two complex linear equations for the infinitesimal variation of the qubit spinor contains 6 parameters: three of quantum boosts $\vec{a}$ and three of quantum rotations $\vec{\omega}$. Notice, due to  (\ref{23}) the left part of (\ref{26}) is proportional to the qubit spinor with the complex multiplier
\begin{eqnarray}\label{34}
C=-
(\Gamma^m_{mn} \Phi_{\beta}^n+\frac{\partial
\Phi_{\beta}^n}{\partial \pi^n}) \Omega^{\beta} .
\end{eqnarray}
Let me write this equations follows
\begin{eqnarray}\label{35}
  \delta \eta &=& \left( \begin {array}{cc} -\frac{i}{2} ( \omega_3+ia_3 )
&-\frac{i}{2} ( \omega_1+ia_1 -i ( \omega_2+ia_2)) \cr
-\frac{i}{2}
 ( \omega_1+ia_1+i ( \omega_2+ia_2))
 &-\frac{i}{2} ( -\omega_3-ia_3)
\end {array} \right) \left(
  \begin{array}{cc}
    \frac{<N|\Omega^{\alpha} \hat{\lambda}_{\alpha}|\Psi>}{<N|N>}  \cr
    \frac{<\Psi|\Omega^{\alpha} \hat{\lambda}_{\alpha}|\Psi>}{<\Psi|\Psi>}
    \end{array}
\right) \cr
    &=& C  \left(
    \begin{array}{cc}
    \frac{<N|\Omega^{\alpha} \hat{\lambda}_{\alpha}|\Psi>}{<N|N>}  \cr
    \frac{<\Psi|\Omega^{\alpha} \hat{\lambda}_{\alpha}|\Psi>}{<\Psi|\Psi>}
  \end{array}
\right).
\end{eqnarray}
One has therefore the eigen-problem and it is easy to find that
\begin{eqnarray}\label{36}
C=-
(\Gamma^m_{mn} \Phi_{\beta}^n+\frac{\partial
\Phi_{\beta}^n}{\partial \pi^n}) \Omega^{\beta}= \pm \frac{1}{2}\sqrt{\vec{a}^2-\vec{\omega}^2 - 2i(\vec{a}\vec{\omega})}.
\end{eqnarray}
Eigen-vectors for these complex eigen-values are as follows:
\begin{eqnarray}\label{37}
   V_1=\left( \begin {array}{cc} -\frac{i(\omega_1+ia_1-i\omega_2+a_2)}{\sqrt{\vec{a}^2-\vec{\omega}^2 - 2i(\vec{a}\vec{\omega})}+i\omega_3-a_3}  \cr
1
\end {array} \right), \quad V_2=\left( \begin {array}{cc} 1 \cr
-\frac{i(\omega_1+ia_1-i\omega_2+a_2)}{-\sqrt{\vec{a}^2-\vec{\omega}^2 - 2i(\vec{a}\vec{\omega})}+i\omega_3-a_3}
\end {array} \right).
\end{eqnarray}
I put $A=\vec{a}^2-\vec{\omega}^2 $ and $B=- 2(\vec{a}\vec{\omega})$, then $4C^2=A+iB$ and four equations arising under square from eigen-problem
\begin{eqnarray}\label{38}
   \left( \begin {array}{cc} -\frac{i}{2} ( \omega_3+ia_3 )
&-\frac{i}{2} ( \omega_1+ia_1 -i ( \omega_2+ia_2)) \cr
-\frac{i}{2}
 ( \omega_1+ia_1+i ( \omega_2+ia_2))
 &-\frac{i}{2} ( -\omega_3-ia_3)
\end {array} \right) \left(
  \begin{array}{cc}
    u_1+iv_1 \\
    u_2+iv_2 \end{array}
\right) \cr
    = C  \left(
    \begin{array}{cc}
    u_1+iv_1  \\
    u_2+iv_2
  \end{array}
\right).
\end{eqnarray}
Thereby, one has two complex (four real) equations for six variables
$\vec{\omega}, \vec{a}$ as functions of four real variables $u_1, v_1, u_2, v_2$ and two equations of two real $A,B$. Analytical solution of this system is not found.

\section{Energy distribution in the ``lump"}
It is clear that spatially non-local description of the extended state could not be relativistically invariant. It means that two vertices say, A and B, arising under interaction of some detecting field with spatially extended soliton-like object lie outside of mutual light cones. In this case quantum measurement at A (in the frame moving toward A) determines value of some dynamical variable at B and vice verse.
A long time this fact was the main argument against non-local quantum field theory. However there are no natural reasons for requirements of conservation of the \textbf{global} causal relations and space-time locality in self-interacting extended systems. For such kind of quantum systems, the relativity should be accompanied by super-relativity to the choice of functional reference frame \cite{Le1,Le2,Le3}.  Namely, broken Lorentz symmetry widely discussed now (see, say, \cite{Kostel}), should be locally restored with help the affine parallel transport  of the local Hamiltonian in the projective Hilbert state space that leads to extended soliton-like solutions \cite{Le2}. It is defined by the velocity of variation of qubit spinor $\eta$ during parallel transport of local Hamiltonian. Moreover, there is some affine gauge field which in some sense restores global Lorentz invariance since the filed equations (23) for for the lump are relativistically invariant.  In fact not any classical field in space-time correspond to the parallel transport in $CP(N-1)$, but in dynamical space-time permissible only fields corresponding to conservation laws in $CP(N-1)$. These conservation laws are expressed by the affine parallel transport. The parallel transport of the local Hamiltonian provides the ``self-conservation" of extended object, i.e. the affine gauge fields couple the soliton-like system (\ref{40}) discussed in \cite{Le1,Le3}.

The field equations for the $SU(N)$ parameters $\Omega^{\alpha}$
dictated by the affine parallel transport of the Hamiltonian vector field
$H^i=\hbar \Omega^{\alpha}\Phi^i_{\alpha}$
\begin{eqnarray}\label{39}
\frac{\delta H^k}{\delta \tau} &= &\hbar \frac{\delta
(\Phi^k_{\alpha} \Omega^{\alpha})}{\delta \tau}=0,
\end{eqnarray}
are quasi-linear PDE
\begin{equation}\label{40}
\frac{\delta \Omega^{\alpha}}{\delta \tau} = V^{\mu} \frac{\partial \Omega^{\alpha}}{\partial x^{\mu} } = -
(\Gamma^m_{mn} \Phi_{\beta}^n+\frac{\partial
\Phi_{\beta}^n}{\partial \pi^n}) \Omega^{\alpha}\Omega^{\beta},
\quad \frac{d\pi^k}{d\tau}= \Phi_{\beta}^k \Omega^{\beta}.
\end{equation}
These field equations describes energy distribution in the lump which does not exist a priori but is becoming during the self-interaction.  The PDE equation obtained as a consequence of the parallel transport of the local Hamiltonian for two-level system living in $CP(1)$ has been shortly discussed \cite{Le4,LeHor}
\begin{equation}\label{41}
\frac{r}{c}\psi_t+ct\psi_r=F(u,v) \rho \cos \psi.
\end{equation}
The one of the exact solutions of this quasi-linear PDE is
\begin{eqnarray}\label{42}
\psi_{exact}(t,r)= \arctan \frac{\exp(2c\rho F(u,v)
f(r^2-c^2t^2))(ct+r)^{2F(u,v)}-1}{\exp(2c\rho F(u,v)
f(r^2-c^2t^2))(ct+r)^{2F(u,v)}+1},
\end{eqnarray}
where $f(r^2-c^2t^2)$ is an arbitrary function of the interval.
It is interesting that this non-monotonic distribution of the force field
describing ``lump" \cite{Le1,Le2,Le3,Le4,Le5,LeHor} that looks like a bubble in the dynamical space-time. The question about stability of this solution and whether such approach deletes the necessity of some additional stabilization forces should be studied carefully.

\section{Summary}
Affine gauge field associated with parallel transport of the local Hamiltonian in $CP(N-1)$ compensates the breakdown of Lorentz symmetry arose due to non-locality of such ``elementary" particle. Phenomenologically it may appear as new states or particles resulting ``deformation" of the Hamiltonian during the parallel transport and the continuous ``measurement" provided by ``quantum boosts" and ``quantum rotations" of local Lorentz reference frame. In other words: in order to avoid the contradiction with causality, the local Lorentz reference frame should be adapted during ``scanning along lump". Such local Lorentz reference frame has been built above whose ``quantum boosts" and ``quantum rotations" are defined by formulas (\ref{36}) and (\ref{38}). Here we have example of relativistic non-local solution (lump) arose due to restoration of the Lorentz symmetry. Extended lumps represent smooth transition from one GCS to another. The conservation laws of LDV's lead to non-monotonic distribution of the force field describing ``lump" that looks like a bubble in the dynamical space-time. Probably it deletes the necessity of some additional stabilization forces preventing flying apart the ``elementary" particle.

Objective quantum measurement is understood here as a comparison of the local dynamical
variables and the dynamical space-time serves as ``measurement machine" for the measurement
encoding. Summarizing it is possible to say that there is an unification of such kind
of quantum measurement and dynamical space-time structure, namely: space-time does not exists as a physical entity without self-interacting quantum ``lump" used as a ``pointer".

Therefore, the geometric formulation of QM being taken not as embellishment but as serious reconstruction, paves the way to new physical interpretation resolving old paradoxes (EPR, Schr\"odinger's Cat), namely: standard QM is incomplete and non-local \cite{Le3}. It requires reformulation in accordance with super-relativity like the classical mechanics was reformulated in accordance with Lorentz invariance of Maxwell equations.

\end{document}